# Analytical Solution of Transverse Oscillation in Cyclotron Using LP Method

Kai Zhou[1]（周凯）, Yun-Tao Song[2]（宋云涛）, Kai-Zhong Ding[2]（丁开忠）, Jian Ge[2]（葛剑）, Kai Yao[3]（姚凯）

[1]*School of Physics, University of Science and Technology of China, Hefei230031, Anhui, China*
[2]*Institute of Plasma Physics Chinese Academy of Sciences, Hefei230031, Anhui, China*
[3]*Hefei CAS Ion Medical and Technical Devices Co., Ltd, Hefei230031, Anhui, China*

**Abstract:** We have carried out an approximate analytical solution to precisely consider the influence of magnetic field on the transverse oscillation of particles in cyclotron. The differential equations of transverse oscillation are solved from the Lindstedt-Poincare method. After careful deduction, the accurate first order analytic solutions are obtained. The analytical solutions are applied to the magnetic field, comes from an isochronous cyclotron with four spiral sectors, the accuracy of these analytical solutions is verified and confirmed from the comparison of numerical method. Finally, we discussed the transverse oscillation at $v_0 = \dfrac{N}{2}$, using the same analytical solution.

**Keywords:** cyclotron; perturbation method; transverse oscillation; LP method

**PACS:** 29.20.-c, 29.27.-a

## 1. Introduction

The motion of particles in cyclotron can be precisely investigated from numerical method, but the magnetic field effect on particle motion can not be visualized. An analytical formula with certain accuracy can consider and visualize the effect of magnetic field.

There are many analytical formulas for transverse oscillation which describe synchrotron and classic cyclotron at relatively simple magnetic field, as shown in Ref.[1]. However, no one has reported such formulas which characterize cyclotron with spiral sectors.

Formulas for calculating transverse oscillation frequency have been reported by several investigators, as shown in Ref.[2][3]. However, the accuracy of these formulas is limited.

In this paper, we solve the linear equations for transverse oscillation through LP method. We have obtained different formulas for calculating transverse oscillation and oscillation frequency. Finally, these formulas were compared with numerical results which are proved to be well accurate.

## 2. Linear equations for transverse oscillation

In Ref.[4], the linear equations for transverse oscillation around the equilibrium orbit are given as:

$$\begin{cases} \dfrac{dx}{d\theta} = \dfrac{p_{re}}{p_{\theta e}} \cdot x + \dfrac{r_e \cdot p^2}{p_{\theta e}^3} \cdot p_x \\ \dfrac{dp_x}{d\theta} = -\dfrac{e}{P} \dfrac{\partial rB_z}{\partial r}\bigg|_{seo} \cdot x - \dfrac{p_{re}}{p_{\theta e}} \cdot p_x, \end{cases} \quad (1)$$

$$\begin{cases} \dfrac{dz}{d\theta} = \dfrac{r_e}{p_{re}} \cdot p_z \\ \dfrac{dp_z}{d\theta} = \dfrac{e}{P}\left(r\dfrac{\partial B_z}{\partial r} - \dfrac{p_r}{p_\theta}\dfrac{\partial B_z}{\partial \theta}\right)\Big|_{seo} x. \end{cases} \quad (2)$$

Where (1) is for radial oscillation and (2) is for axial oscillation, x and z are respectively radial and axial offset of the equilibrium orbit, $p_x$ and $p_z$ are correspondingly radial and axial momentum offset of the equilibrium orbit.

The coefficients of the above linear differential equations are determined from the parameters of equilibrium orbit such as $r_e$ and $p_{re}$. As long as the magnetic fields are given, the equilibrium orbit can be determined from numerical integration or Gordon's formulas [3] followed by the determination of the coefficients of the equation.

In order to simplify these equations, let

$$\begin{cases} a(\theta) = \dfrac{p_{re}}{p_{\theta e}}, \quad b(\theta) = \dfrac{r_e}{p_{\theta e}^3}, \\ c(\theta) = -\dfrac{e}{P}\dfrac{\partial rB_z(r,\theta)}{\partial r}\Big|_{seo}, \quad d(\theta) = -\dfrac{p_{re}}{p_{\theta e}}, \\ e(\theta) = \dfrac{r_e}{p_{re}}, \quad f(\theta) = \dfrac{e}{P}\left(r\dfrac{\partial B_z}{\partial r} - \dfrac{p_r}{p_\theta}\dfrac{\partial B_z}{\partial \theta}\right)\Big|_{seo}. \end{cases} \quad (3)$$

Then the radial oscillation equations can be written as

$$\begin{cases} \dfrac{dx}{d\theta} = a(\theta)\cdot x + b(\theta)\cdot p_x \\ \dfrac{dp_x}{d\theta} = c(\theta)\cdot x + d(\theta)\cdot p_x. \end{cases} \quad (4)$$

By using the following formula:

$$x(\theta) = \sqrt{b(\theta)} \cdot u(\theta), \quad (5)$$

(4) can be rewritten as a Hill's equation,

$$u'' + G(\theta)\cdot u = 0, \quad (6)$$

where

$$G(\theta) = -\dfrac{3}{4}\left(\dfrac{b'}{b}\right)^2 + \dfrac{1}{2}\dfrac{b''}{b} + (ad - bc) + a\dfrac{b'}{b} - a'. \quad (7)$$

Here $G(\theta)$ has the same periodicity as the equilibrium orbit. Expand $G(\theta)$ into Fourier series:

$$G(\theta) = v^2 + \sum_n P_n \cos n\theta + Q_n \sin n\theta. \quad (8)$$

In order to simplify, let

$$g(\theta) = \sum_n P_n \cos n\theta + Q_n \sin n\theta. \quad (9)$$

Then the equation (6) becomes

$$u'' + v^2 u = -g(\theta)u. \quad (10)$$

(10) is a simplified radial oscillation equation. We can get directly the simplified axial equation by replacing (5) and (7) with the following two formulas:

$$z(\theta) = \sqrt{e(\theta)} \cdot u(\theta), \tag{11}$$

$$G(\theta) = \frac{1}{2}\frac{e''}{e} - \frac{3}{4}\left(\frac{e'}{e}\right)^2 - ef. \tag{12}$$

## 3. Solving oscillation equations by LP method

Equation (10) is a linear differential equation with variable coefficients, and the LP method is very effective for solving such equation.

Now, we want to solve equation (10) by using LP method. First, we introduce a small parameter $\varepsilon$ with the condition of $0 \leq \varepsilon \leq 1$ and rewrite (10) as follows:

$$u'' + v^2 u = -\varepsilon \cdot g(\theta) u. \tag{13}$$

According to LP method, we will try to solve (13) by inserting a perturbation series:

$$u(\theta) = u_0(\theta) + \varepsilon \cdot u_1(\theta) + \varepsilon^2 \cdot u_2(\theta) + \cdots, \tag{14}$$

$$v^2 = v_0^2 + \varepsilon \cdot v_1^2 + \varepsilon^2 \cdot v_2^2 + \cdots, \tag{15}$$

and then find the expressions of $u_0(\theta)$, $u_1(\theta)$, $u_2(\theta)$ as well as the values of $v_0^2$, $v_1^2$, $v_2^2$. Finally, the solutions of (10) is obtained by letting $\varepsilon = 1$.

Substitute (14) and (15) into (13), we can get

$$\begin{aligned}&(u_0'' + \varepsilon \cdot u_1'' + \varepsilon^2 \cdot u_2'' + \cdots) + (v_0^2 + \varepsilon \cdot v_1^2 + \varepsilon^2 \cdot v_2^2 + \cdots)(u_0 + \varepsilon \cdot u_1 + \varepsilon^2 \cdot u_2 + \cdots) \\ &= \varepsilon \cdot g(\theta)(u_0 + \varepsilon \cdot u_1 + \varepsilon^2 \cdot u_2 + \cdots)\end{aligned} \tag{16}$$

The coefficient of each power of $\varepsilon$ should be zero, hence we can get:

$$\begin{cases} u_0'' + v_0^2 \cdot u_0 = 0 \\ u_1'' + v_0^2 \cdot u_1 = -v_1^2 \cdot u_0 - g(\theta) \cdot u_0 \\ u_2'' + v_0^2 \cdot u_2 = -v_2^2 \cdot u_0 - v_1^2 \cdot u_1 - g(\theta) \cdot u_1. \end{cases} \tag{17}$$

From the above procedures, we transform equation (13) into several ordinary differential equations. The solution of the first equation of (17) is

$$u_0(\theta) = A\cos v_0\theta + B\sin v_0\theta. \tag{18}$$

Substitute (18) into the second equation of (17) which becomes as:

$$\begin{aligned}u_1'' + v_0^2 \cdot u_1 &= -v_1^2(A\cos v_0\theta + B\sin v_0\theta) - \sum_n P_n A\cos n\theta \cos v_0\theta + P_n B\cos n\theta \sin v_0\theta \\ &\quad + Q_n A\sin n\theta \cos v_0\theta + Q_n B\sin n\theta \sin v_0\theta,\end{aligned} \tag{19}$$

where $-v_1^2(A\cos v_0\theta + B\sin v_0\theta)$ in the right side is secular term. LP method requires to avoid the secular term by choosing the value of $v_1^2$. It is obvious that $v_1^2$ must be zero to eliminate the secular term, which is

$$v_1^2 = 0. \tag{20}$$

The solution of (19) is:

$$u_1 = \sum_n -\frac{P_n A}{2}\left[\frac{\cos(n+v_0)\theta}{v_0^2-(n+v_0)^2}+\frac{\cos(n-v_0)\theta}{v_0^2-(n-v_0)^2}\right]-\frac{P_n B}{2}\left[\frac{\sin(n+v_0)\theta}{v_0^2-(n+v_0)^2}-\frac{\sin(n-v_0)\theta}{v_0^2-(n-v_0)^2}\right]$$
$$-\frac{Q_n A}{2}\left[\frac{\sin(n+v_0)\theta}{v_0^2-(n+v_0)^2}+\frac{\sin(n-v_0)\theta}{v_0^2-(n-v_0)^2}\right]+\frac{Q_n B}{2}\left[\frac{\cos(n+v_0)\theta}{v_0^2-(n+v_0)^2}-\frac{\cos(n-v_0)\theta}{v_0^2-(n-v_0)^2}\right]. \tag{21}$$

In order to simplify, let

$$\begin{cases} C_1 = v_0^2 - (n+v_0)^2 \\ C_2 = v_0^2 - (n-v_0)^2, \end{cases} \tag{22}$$

substitute (21) into third formula of (17), so the secular term can be written as

$$-v_2^2(A\cos v_0\theta + B\sin v_0\theta) + \sum_n \frac{P_n^2 A}{4}\left(\frac{\cos v_0\theta}{C_1}+\frac{\cos v_0\theta}{C_2}\right)+\frac{P_n^2 A}{4}\left(\frac{\sin v_0\theta}{C_1}+\frac{\sin v_0\theta}{C_2}\right)$$
$$+\frac{Q_n P_n A}{4}\left(\frac{\sin v_0\theta}{C_1}-\frac{\sin v_0\theta}{C_2}\right)-\frac{Q_n P_n B}{4}\left(\frac{\cos v_0\theta}{C_1}-\frac{\cos v_0\theta}{C_2}\right)+\frac{Q_n P_n A}{4}\left(-\frac{\sin v_0\theta}{C_1}+\frac{\sin v_0\theta}{C_2}\right) \tag{23}$$
$$+\frac{Q_n P_n B}{4}\left(\frac{\cos v_0\theta}{C_1}-\frac{\cos v_0\theta}{C_2}\right)+\frac{Q_n^2 A}{2}\left(\frac{\cos v_0\theta}{C_1}+\frac{\cos v_0\theta}{C_2}\right)+\frac{Q_n^2 B}{2}\left(\frac{\sin v_0\theta}{C_1}+\frac{\sin v_0\theta}{C_2}\right).$$

The secular term can be eliminated by choosing the value of $v_2^2$. Let (23) is equal to zero, then we can get

$$v_2^2 = \sum_n \frac{P_n^2 + Q_n^2}{4}\left[\frac{1}{v_0^2-(n+v_0)^2}+\frac{1}{v_0^2-(n-v_0)^2}\right]. \tag{24}$$

Substitute (20) and (24) into (15), and let $\varepsilon = 1$, equation (15) becomes as

$$v^2 = v_0^2 + \sum_n \frac{P_n^2 + Q_n^2}{4}\left[\frac{1}{v_0^2-(n+v_0)^2}+\frac{1}{v_0^2-(n-v_0)^2}\right]. \tag{25}$$

Since the value of $v^2$, $P_n$ and $Q_n$ have been determined from (8), so, we can get the value of $v_0$ by solving the above equation with the help of bisection method.

So far, we have found the expressions of $u_0$, $u_1$ and the values of $v_0, v_1, v_2$. By letting $\varepsilon = 1$, eventually we can get the solution of (10):

$$u = u_0 + u_1$$
$$= A\cos v_0\theta + B\sin v_0\theta$$
$$+\sum_n -\frac{P_n A}{2}\left[\frac{\cos(n+v_0)\theta}{v_0^2-(n+v_0)^2}+\frac{\cos(n-v_0)\theta}{v_0^2-(n-v_0)^2}\right]-\frac{P_n B}{2}\left[\frac{\sin(n+v_0)\theta}{v_0^2-(n+v_0)^2}-\frac{\sin(n-v_0)\theta}{v_0^2-(n-v_0)^2}\right] \tag{26}$$
$$-\frac{Q_n A}{2}\left[\frac{\sin(n+v_0)\theta}{v_0^2-(n+v_0)^2}+\frac{\sin(n-v_0)\theta}{v_0^2-(n-v_0)^2}\right]+\frac{Q_n B}{2}\left[\frac{\cos(n+v_0)\theta}{v_0^2-(n+v_0)^2}-\frac{\cos(n-v_0)\theta}{v_0^2-(n-v_0)^2}\right],$$

where A and B can be determined from the initial conditions of $u(0)$ and $u'(0)$. Finally, according to the formula of (5), we can get the expression of $x(\theta)$

$$x(\theta) = \sqrt{b(\theta)} \cdot u(\theta). \tag{27}$$

The solution of axial oscillation $z(\theta)$ can be solved from the same method, but the specific process can no longer be repeated here.

## 4. Comparison with numerical solution

The analytical solutions are applied to the magnetic field, which comes from SC200 cyclotron. The SC200 cyclotron is a compact superconducting proton cyclotron with four spiral sectors, used for proton therapy, the last energy of SC200 is 200 MeV, the extracting radius is 60cm, and the central magnetic field is 2.95T. The transverse oscillation under the initial condition of $E_k = 100 MeV$, $x_0 = 2mm$, $x'_0 = 0$, $z_0 = 2mm$, $z'_0 = 0$ was calculated by (25), (26) and (27). Further more, we calculated the equilibrium orbit and the general orbit by solving the equation of motion using the 4th order Runge-Kutta method, with the above initial condition. The equation of motion and the numerical method of calculating the equilibrium orbit comes from Ref.[9], then the transverse oscillation is obtained by comparing the general orbit with the equilibrium orbit.
And the comparison of the result from the formulas with those of the above numerical method is shown in the following figures.

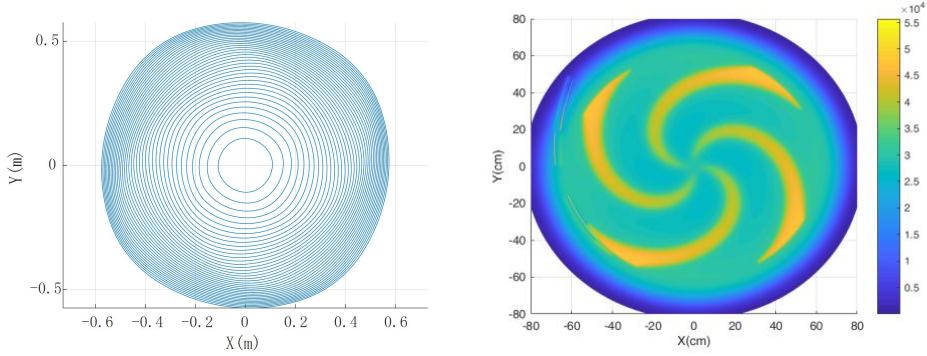

Fig 1. The plot at the left shows the 5~180 MeV equilibrium orbits of SC200, with the energy step of 5 MeV. The plot at the right shows the magnetic field in mid-plane.

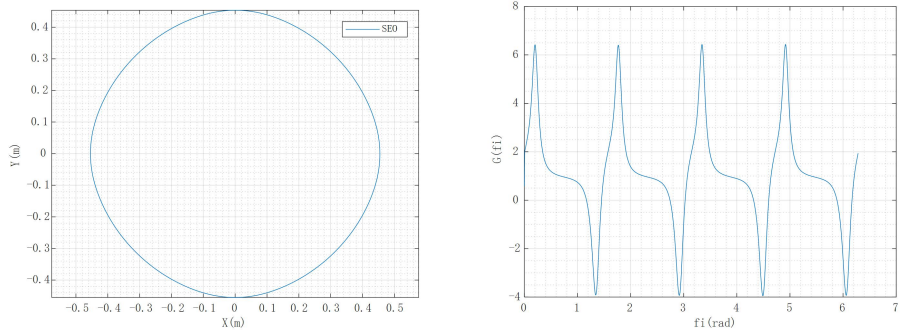

Fig 2. The plot at the left-hand side shows the equilibrium orbit for a 100 MeV particle whereas the plot at the right-hand side shows $G(\theta)$, obtained from (7).

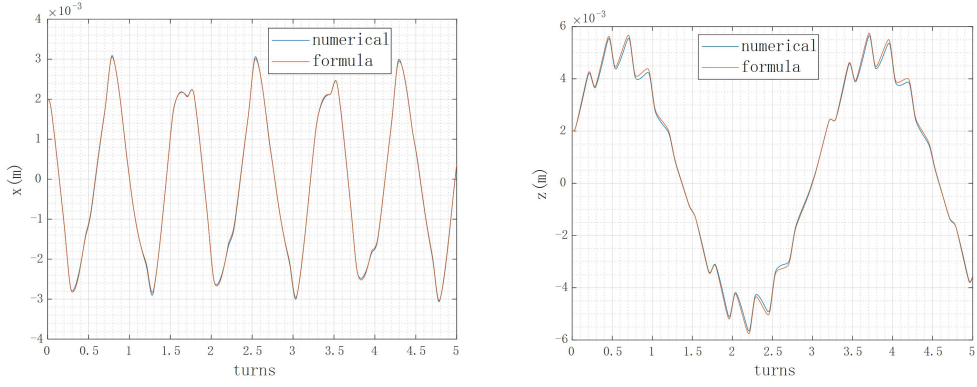

Fig 3. The oscillation around the equilibrium orbit, obtained from the formula (26), (27) (red curve), and that of numerical method (blue curve) which are superimposed. The plot at the left shows the radial oscillation around the 100 MeV equilibrium orbit. While the plot at the right-hand side shows the axial oscillation around the 100 MeV equilibrium orbit. As we can see, the differences are almost indistinguishable.

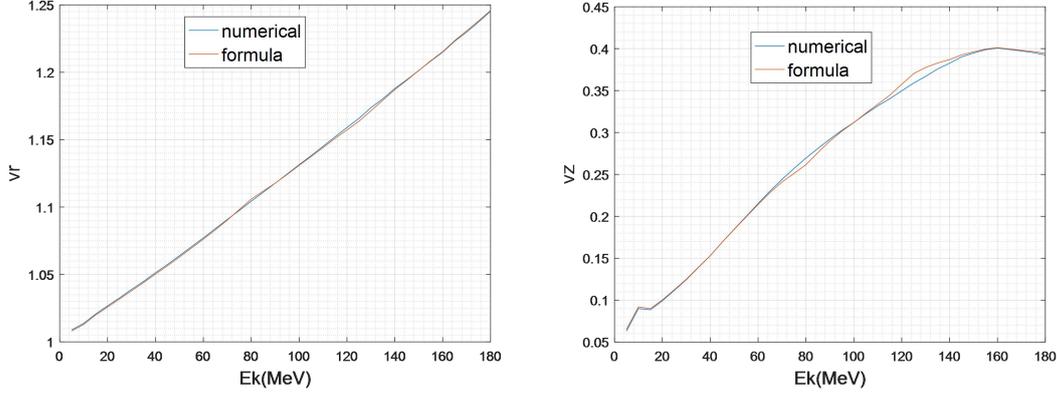

Fig 4. The oscillation frequency $v_0$, obtained from formula (25) (red curve) and numerical method (blue curve) which are superimposed. The plot at the left-hand side shows the radial frequency $v_r$ and the plot at the right shows the axial frequency $v_z$. As can be seen, the difference in radial frequency is negligible. The difference in axial frequency is also very small except at the energy of 80 MeV and 130 MeV. The deviation at 80 MeV and 130 MeV may be due to the unavoidable error.

## 5. Improved formulas when $v_0 = \dfrac{N}{2}$

The value of the denominator $v_0^2 - (n - v_0)^2$ in (26) will be zero at $v_0 = \dfrac{N}{2}$, where $n = 1, 2, 3 \cdots$ and $N$ is an integer. So, the formula (26) will be invalid at $v_0 = \dfrac{N}{2}$.

Therefore, further processing of equation (26) is required. Actually, equation (26) is deduced based on the initial condition such as

$$\begin{cases} u(0) = u_0(0) + u_1(0) + u_2(0) + \cdots, \\ u'(0) = u_0'(0) + u_1'(0) + u_2'(0) + \cdots. \end{cases} \quad (29)$$

Now, we solve equation (10) again based on a new initial condition:

$$\begin{cases} u(0) = u_0(0), & u_1(0) = 0, & u_2(0) = 0, \cdots \\ u'(0) = u_0'(0), & u_1'(0) = 0, & u_2'(0) = 0, \cdots \end{cases} \tag{30}$$

then $u(\theta)$ is obtained as:

$$u(\theta) = (A + A')\cos v_0\theta + (B + B')\sin v_0\theta + \sum_n -\frac{P_n A}{2}\left[\frac{\cos(n+v_0)\theta}{v_0^2 - (n+v_0)^2} + \frac{\cos(n-v_0)\theta}{v_0^2 - (n-v_0)^2}\right]$$

$$-\frac{P_n B}{2}\left[\frac{\sin(n+v_0)\theta}{v_0^2 - (n+v_0)^2} - \frac{\sin(n-v_0)\theta}{v_0^2 - (n-v_0)^2}\right] - \frac{Q_n A}{2}\left[\frac{\sin(n+v_0)\theta}{v_0^2 - (n+v_0)^2} + \frac{\sin(n-v_0)\theta}{v_0^2 - (n-v_0)^2}\right] \tag{31}$$

$$+\frac{Q_n B}{2}\left[\frac{\cos(n+v_0)\theta}{v_0^2 - (n+v_0)^2} - \frac{\cos(n-v_0)\theta}{v_0^2 - (n-v_0)^2}\right],$$

where:

$$\begin{cases} A = u(0), \quad A' = \sum_n \frac{P_n A}{2}\left[\frac{1}{C_1} + \frac{1}{C_2}\right] - \frac{Q_n B}{2}\left[\frac{1}{C_1} - \frac{1}{C_2}\right], \\ B = \frac{u'(0)}{v_0}, \quad B' = \frac{\sum_n \frac{P_n B}{2}\left[\frac{n+v_0}{C_1} - \frac{n-v_0}{C_2}\right] + \frac{Q_n A}{2}\left[\frac{n+v_0}{C_1} + \frac{n-v_0}{C_2}\right]}{v_0}. \end{cases} \tag{32}$$

The initial condition (29) is equivalent to (30), while the difference between the resulting formulas such as (26) and (31) is a small quantity of high order. In contrast, formula (26) has high accuracy, but the preference of (31) is due to the constant coefficients of A and B during $v_0 \to \frac{N}{2}$, which encourage further simplification.

So, make the following transformation for the formulas in square brackets of (31), such as the first square bracket

$$\frac{\cos(n+v_0)\theta}{v_0^2 - (n+v_0)^2} + \frac{\cos(n-v_0)\theta}{v_0^2 - (n-v_0)^2}$$

$$= \frac{\cos(n+v_0)\theta}{v_0^2 - (n+v_0)^2} + \frac{\cos(n-v_0)\theta}{v_0^2 - (n-v_0)^2} - \frac{\cos v_0\theta}{v_0^2 - (n-v_0)^2} + \frac{\cos v_0\theta}{v_0^2 - (n-v_0)^2} \tag{33}$$

$$= \frac{\cos(n+v_0)\theta}{v_0^2 - (n+v_0)^2} + \frac{\cos v_0\theta}{v_0^2 - (n-v_0)^2} + \frac{-2\cdot \sin\frac{n}{2}\theta \cdot \sin\frac{n-2v_0}{2}\theta}{n(2v_0 - n)}.$$

After above transformation, the $u(\theta)$ becomes as

$$u(\theta) = u_0(\theta) + u_1(\theta)$$
$$= A^* \cos v_0\theta + B^* \sin v_0\theta + \sum_n C_n^* \cdot \cos(n+v_0)\theta + D_n^* \cdot \sin(n+v_0)\theta$$
$$+ \sum_n \frac{P_n A}{C_2} \sin\frac{n}{2}\theta \sin\frac{n-2v_0}{2}\theta - \frac{P_n B}{C_2}\sin\frac{2v_0-n}{2}\theta\cos\frac{n}{2}\theta \tag{34}$$
$$- \frac{Q_n A}{C_2}\cos\frac{n}{2}\theta\sin\frac{n-2v_0}{2}\theta + \frac{Q_n B}{C_2}\sin\frac{n}{2}\theta\sin\frac{n-2v_0}{2}\theta,$$

where:

$$\begin{cases} A^* = A + A' + \sum_n -\frac{P_n A}{2}\frac{1}{C_2} - \frac{Q_n B}{2}\frac{1}{C_2} \\ B^* = B + B' + \sum_n \frac{P_n B}{2}\frac{1}{C_2} - \frac{Q_n A}{2}\frac{1}{C_2} \\ C_n^* = \left(\frac{-P_n A}{2} + \frac{Q_n B}{2}\right)\frac{1}{C_1} \\ D_n^* = \left(\frac{-P_n B}{2} - \frac{Q_n A}{2}\right)\frac{1}{C_1}. \end{cases} \quad (35)$$

It is easy to find that $C_n^*$ and $D_n^*$ remain constant during $v_0 \to \frac{N}{2}$, according to the initial condition $A^*$ and $B^*$ can also be written as:

$$\begin{cases} A^* = u(0) - \sum_n \left(\frac{-P_n A}{2} + \frac{Q_n B}{2}\right)\frac{1}{v_0^2 - (n+v_0)^2} \\ B^* = \dfrac{u'(0) + \sum_n \dfrac{P_n B + Q_n A}{2}\dfrac{n+v_0}{v_0^2 - (n+v_0)^2} + \dfrac{P_n B - Q_n A}{2n}}{v_0}, \end{cases} \quad (36)$$

which indicate that $A^*$ and $B^*$ also remain constant during $v_0 \to \frac{N}{2}$.

So, the expression of $u(\theta)$ under $v_0 = \frac{N}{2}$ can be obtained as

$$\begin{aligned} \lim_{v_0 \to \frac{N}{2}} u(\theta) =\ & A^* \cos\frac{N}{2}\theta + B^* \sin\frac{N}{2}\theta + \sum_n C_n^* \cos\left(n + \frac{N}{2}\right)\theta + D_n^* \sin\left(n + \frac{N}{2}\right)\theta \\ & + \sum_{n \neq N} \frac{P_n A}{C_2}\sin\frac{n}{2}\theta \sin\frac{n-N}{2}\theta - \frac{P_n B}{C_2}\sin\frac{N-n}{2}\theta \cos\frac{n}{2}\theta \\ & - \frac{Q_n A}{C_2}\cos\frac{n}{2}\theta \sin\frac{n-N}{2}\theta + \frac{Q_n B}{C_2}\sin\frac{n}{2}\theta \sin\frac{n-N}{2}\theta \\ & - \frac{P_N A}{2N}\cdot\theta\cdot\sin\frac{N}{2}\theta - \frac{P_N B}{2N}\cdot\theta\cdot\cos\frac{N}{2}\theta - \frac{Q_N A}{2N}\cdot\theta\cdot\cos\frac{N}{2}\theta - \frac{Q_N B}{2N}\cdot\theta\cdot\sin\frac{N}{2}\theta. \end{aligned} \quad (37)$$

(37) is an improved formula for calculating oscillation at $v_0 = \frac{N}{2}$. As we can see, formula (37) contains the term of $\theta\cdot\sin\frac{N}{2}\theta$ and $\theta\cdot\cos\frac{N}{2}\theta$, which indicates that the amplitude of $u(\theta)$ increases with respect of azimuth.

Here we have calculated the transverse oscillation at $v_0 = 1$ for two different particles with the initial condition of $A = 0.02$, $B = 0.04$ and $A = 0.006$, $B = -0.04$ using the formula (37). The values of $P_n, Q_n$ comes from the parameters of 2.7 MeV equilibrium orbit, where the corresponding $v_0$ is very close to 1. The result is shown as following:

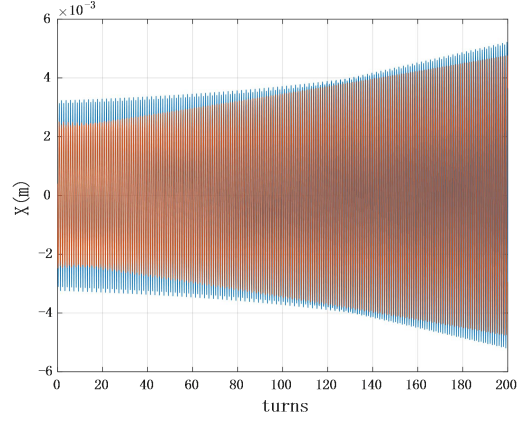

Fig 5. Calculated results of radial oscillation from formulas (37); As can be seen, the amplitude of oscillation increases with respect to azimuth at $v_0 = 1$.

## 6. Summary

In this work, we have proposed approximate formulas for the calculation of transverse oscillation, around the equilibrium orbit, using LP method. These formulas were further confirmed from the nice correlation of numerical and LP based results. Moreover, these formulas are highly accurate for the calculation of transverse oscillation and the oscillation frequency. Finally, we discussed the transverse oscillation at $v_0 = \dfrac{N}{2}$, where the amplitude increases with azimuth.

The numerical method can be used to study the transverse oscillation precisely, but there are some limitations. For example, firstly it is unavailable to directly demonstrate the influence of magnetic field structures on motions of particles; secondly, in a given magnetic field with fixed working diagrams, particles may not cross the resonance lines that are proposed to be studied, or they have passed it but the influence imposed on motions of particles is not obvious and thus no sufficient information can be obtained. On the contrary, analytical solution seems more flexible. The transverse oscillation of particles under any resonance line can be studied by only changing the parameters of the magnetic field in the formulas concerned while with no necessity to change the whole magnetic field, and the influence of the Nth harmonic on the oscillation oscillation can be studied by changing the amplitude of the Nth harmonic in the given magnetic field.

The method of calculating transverse oscillation given in this paper can also be applied in the transverse oscillation equation with high order terms and coupling terms, more information about the high-order resonance and coupling resonance may be obtained.